\documentclass[aps,reprint,nofootinbib,prl]{revtex4-2}
\usepackage{bvrshorthand}
\usepackage{tcolorbox}
\usepackage{tikz}
\usepackage[hidelinks]{hyperref}
\usetikzlibrary{shapes.misc}
\usetikzlibrary{decorations.markings}
\usetikzlibrary{decorations.pathmorphing, patterns,shapes}
\usetikzlibrary{calc,shapes.geometric}

\tikzset{cross/.style={cross out, draw=black, minimum size=2*(#1-\pgflinewidth), inner sep=0pt, outer sep=0pt},
	cross/.default={1pt}}

\begin{document}
\title{The doubly conformal dispersion relation}

\author{Maddalena Ferragatta and Balt C.~van Rees}

\affiliation{CPHT, CNRS, École Polytechnique, Institut Polytechnique de Paris, 91120 Palaiseau, France}

\begin{abstract}
	We provide a simple derivation of the conformal dispersion relation of \cite{Carmi:2019cub} which reconstructs a CFT four-point function from its double discontinuity. We exhibit a hidden conformal symmetry which explains the appearance of the `magic variable' $x$ as a simple cross-ratio, built from four physical cross-ratios. We also introduce a regulator in the double discontinuity operation and then prove that one subtraction suffices for all integrals to converge.
\end{abstract}

\maketitle
\section{Introduction}
For conformal four-point functions, the correct dispersive object is the double commutator $\vev{[\phi_1, \phi_2] [\phi_3, \phi_4]}$. This was explicated by the ground-breaking Lorentzian inversion formula (LIF) of Caron-Huot \cite{Caron-Huot:2017vep}, which shows that the OPE spectral density $\rho(\Delta,J)$ of \cite{Dobrev:1977qv} can be reconstructed from this double commutator. By combining this formula with the Euclidean partial wave decomposition (EPWD), Carmi and Caron-Huot in \cite{Carmi:2019cub} arrived at the \emph{conformal dispersion relation} (CDR) which reconstructs the correlation function from its double commutator, equation \eqref{originalCDR} below.

The kernel $\cK$ in the CDR is mysterious in several ways. It consists of a `contact' and `bulk' term, and up to some prefactors the latter is just a ${}_2 F_1$ hypergeometric function of a `magical' \cite{Caron-Huot:2020adz} variable $x(u,v,u',v')$. But where does this peculiar structure come from? Secondly, the CDR is purely a result in complex analysis, but current derivations rely on either the EPWD \cite{Carmi:2019cub} or Mellin space \cite{Caron-Huot:2020adz}. But why do we need such harmonic approaches for a result in complex analysis? And how general is the CDR? Does it apply only to functions with a convergent conformal block decomposition? Conversely, is the integration kernel unique or are there alternative dispersion relations?

We consider it especially important to answer these questions because we view the CDR as the most fundamental structure from which one should (logically, not historically) \emph{derive} related structures like the aforementioned LIF, dispersive functionals \cite{Caron-Huot:2020adz,Caron-Huot:2021enk}, dispersion relations in Mellin space \cite{Mack:2009mi,Penedones:2019tng}, and the Polyakov-Regge block decomposition \cite{Mazac:2019shk,Sleight:2019ive,Caron-Huot:2020adz} (see also \cite{Gopakumar:2016wkt,Gopakumar:2018xqi,Polyakov:1974gs} for related earlier ideas). This work should therefore be considered part of a program where all of these are to be proven in a mathematically rigorous way, solidifying our understanding of higher-dimensional CFT correlation functions for future generations of humans and machines.\footnote{Our results also apply beyond CFTs: the flat-space limit of QFT in AdS \cite{Paulos:2016fap} allows us to infer universal properties of non-perturbative QFT scattering amplitudes \cite{vanRees:2022zmr,vanRees:2023fcf,Caron-Huot:2021enk} by viewing them as limits of conformal correlation functions.}

As stated in the abstract, we (humans) have discovered an elegant first-principles derivation of the CDR which relies purely on single-variable complex analysis. The variable $x$ will be a `cross-ratio of cross-ratios' and arises because of a hidden conformal symmetry in the problem.

\section{Original setup}
We consider scalar conformal correlation functions
\begin{equation}
	\cG^\text{phys}(u,v) \colonequals \vev{\phi(p_1) \phi(p_2) \phi(p_3) \phi(p_4)} p_{13}^{2\De} p_{24}^{2\De}
\end{equation}
with the analytic properties stemming from absolutely convergent OPEs with integer spins in all channels, and with $u$, $v$ the usual cross-ratios. The principal region of interest will always be $u,v > 0$ which covers both purely Euclidean and spacelike Lorentzian configurations (separated by the black dashed lines in figure \ref{fig:regionsab} below). The lightcone asymptotics of the physical correlator are then
\begin{align}
	\label{simpleasumptotics}
	\cG^\text{phys}(u,v) &\overset{u \to 0}{\sim} u^{-\De}, \qquad \, \cG^\text{phys}(u,v) \overset{v \to 0}{\sim} v^{-\De},\\ &  \cG^\text{phys}(u,v) \overset{u, v \to \infty}{\sim} 1,\nn
\end{align}
which corresponds to identity operator exchange in the three channels. As discussed in appendix \ref{app:convergence}, the large $u,v$ asymptotics are not quite soft enough, so we need to introduce a \emph{subtracted} correlator. We define it as
\begin{equation}
	\label{subtractionmaintext}
	\cG(u,v) \colonequals \frac{\cG^{\text{phys}}(u,v)}{4 \sqrt{u v}}\,.
\end{equation}
In terms of the cross-ratios double commutators become double discontinuities, which are defined as
\begin{equation}
	\label{dDiscs}
	\dDisc_s[\cG](u,v) \colonequals  \cG(u,v) - \frac{1}{2} \cG(e^{2 \pi i} u, v) - \frac{1}{2} \cG(e^{-2\pi i} u, v)
\end{equation}
and $\dDisc_t[\cG](u,v)$ which is the same operation applied to $v$ instead of $u$. Our goal is now to reconstruct $\cG(u,v)$ entirely from these two double discontinuities.

\section{Modified setup}
We will apply a few modifications to the above problem which will greatly simplify the derivations below.

First, we assume crossing symmetry:
\begin{equation}
	\cG(u,v) = \cG(v,u)\,.
\end{equation}
This can be done without loss of generality. Indeed, if a given $\cG(u,v)$ is not crossing symmetric then it can always be split into a symmetric part and an antisymmetric part, and for the antisymmetric part we can apply the conformal dispersion relation to
\begin{equation}\label{antisymmcorr}
\cG(u,v)/(u - v)
\end{equation}
which is crossing symmetric (and also vanishes sufficiently fast in the Regge limit).

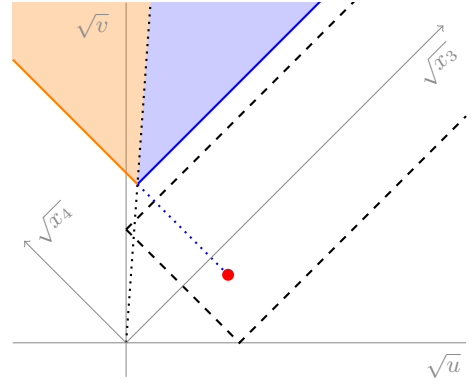
\begin{figure}
	\centering

	\begin{tikzpicture}[scale=1.5] 
    
    \clip (-1, -1) rectangle (3, 3);
    
    \draw[gray!90] (-1, 0) -- (3, 0);
    \draw[gray!90] (0, -0.3) -- (0, 3);
    
    
    \fill[orange, opacity=0.3] 
        (-2, 3.5) -- 
        (-1, 2.5) -- 
        (0.1005, 1.3995) -- 
        (3, 41.776) -- 
        (3, 10) -- 
        (-2, 10) -- cycle;
        
    \fill[blue, opacity=0.2] 
        (0.1005, 1.3995) -- 
        (3, 4.299) -- 
        (3, 41.776) -- cycle;

	\node[gray!90] at (2.8,-0.2) {$\sqrt{u}$};
	\node[gray!90] at (-0.3,2.8) {$\sqrt{v}$};

    
    \draw[black, dashed, thick] (0, 1) -- (1, 0) -- (3, 2); 
    \draw[black, dashed, thick] (0, 1) -- (2, 3) -- (3, 4); 
    
	\draw[blue, thick] (0.1005, 1.3995) -- (3, 4.299);

    \draw[black, dotted, thick] (0, 0) -- (3, 41.776);
    
    
    \draw[orange, thick] (-1, 2.5) -- (0.1005, 1.3995);
    
    \draw[blue, dotted, thick] (0.1005, 1.3995) -- (0.9, 0.6);
    
    \fill[red] (0.9, 0.6) circle (1.5pt);

	\draw[gray!90, ->] (0,0) -- (2.8,2.8);
	\node[gray!90, rotate=45] at (2.7
	5,2.45) {$\sqrt{x_3}$};

	\draw[gray!90, ->] (0,0) -- (-0.9,0.9);
	\node[gray!90, rotate=45] at (-0.65,1.05) {$\sqrt{x_4}$};
\end{tikzpicture}

	\caption{\label{fig:regionsab}The integration domain in the original $(\sqrt{u},\sqrt{v})$ variables. In red the point corresponding to $(x_1, x_2)$. The thick orange line is the integration region for the single-variable dispersion relation of \eqref{contact1}. The blue line is the support of the new term added in \eqref{contact2}. The two-variable integral on the right-hand side of \eqref{contactandbulk} probes $\tilde \cG(x_3, x_4)$ in the blue region and its discontinuity in the orange region. The $\eta$ regulator displaces the $u=0$ branch point of $\tilde \cG(x_3, x_4)$ to the interior of the orange region. Finally, the dashed black lines correspond to the separation of the positive quadrant into the Euclidean domain (where the red point sits) and the three spacelike separated Lorentzian domains. We added these lines to guide the reader, but they play no role in our analysis.}
\end{figure}

Second, we change variables to
\begin{equation}
	\label{xfromuv}
	x_1 = (\sqrt{v} + \sqrt{u})^2, \qquad x_2 = (\sqrt{v} - \sqrt{u})^2\,.
\end{equation}
We explain this change using figure \ref{fig:regionsab}, which shows the $\sqrt{u}$, $\sqrt{v}$ plane, so the original region of interest is still the positive quadrant. The correlator $\cG(u,v)$ has branch cuts starting at the horizontal and vertical axes, and the double discontinuities are defined using discontinuities across these cuts in the regions where either $\sqrt{u}$ or $\sqrt{v}$ (but not both) are negative. The change
\begin{equation}
	\sqrt{x_1} = \sqrt{v} + \sqrt{u}, \qquad \sqrt{x_2} = \sqrt{v} - \sqrt{u}\,,
\end{equation}
now corresponds to a simple rotation in this plane, moving in particular the branch cuts to $\sqrt{x_1} = \pm \sqrt{x_2}$. (In figure \ref{fig:regionsab} the $\sqrt{x_1}$ and $\sqrt{x_2}$ axes are respectively indicated as $\sqrt{x_3}$ and $\sqrt{x_4}$, for later convenience.) The variables $x_1$ and $x_2$ are then simply squares of these variables. Notice that $\sqrt{x_1} > 0$ always so the passage to $x_1$ is one-to-one. On the other hand, $\sqrt{x_2}$ can have either sign but crossing symmetry dictates that the correlator is even in $\sqrt{x_2}$ and so $\cG(u,v)$ has no branch cut at $x_2 = 0$. In fact, for fixed $x_1 > 0$ we know that $x_2 \mapsto \cG(x_1, x_2)$ only has a branch cut at $x_2 \geq x_1$ and is analytic everywhere else.

In terms of $x_1$ and $x_2$ there is only one double discontinuity, which reads
\begin{multline}
	\dDisc[\cG](x_1, x_2) = \cG(x_1, x_2) \\ - \frac{1}{2} \cG(x_2, x_1 + i \epsilon)  - \frac{1}{2} \cG(x_2, x_1 - i \epsilon)\,,
\end{multline}
and which will feature only for real $0 < x_2 \leq x_1$, so in between the axes labeled $\sqrt{x_3}$ and $\sqrt{v}$ in figure \ref{fig:regionsab}. In the new variables, our precise goal is to find a kernel $\cK(x_1, \ldots, x_4)$ such that
\begin{align}
	\label{originalCDR}
	&\cG(x_1, x_2) = \\ & \iint_{0 < x_4 \leq x_3} d x_3 \, d x_4 \, \cK(x_1, x_2, x_3, x_4) \dDisc[\cG](x_3, x_4) \nn
\end{align}
holds for any real $x_1$ and $x_2$ such that $x_1 > 0$ and $x_2 < x_1$.\footnote{It should be straightforward to extend our derivation to any complex $x_2$ away from the cut $x_2 \ge x_1$ by analytic continuation.}

The various one- and two-dimensional integration domains are shown in figure \ref{fig:regionsab}. We will also venture into the complex $x_2$ plane, but $\cG(x_1, x_2)$ is not required to be analytic, or even smooth, in $x_1$. The conditions on $\cG(x_1, x_2)$ that \emph{are} necessary for each step all follow from CFT unitarity, as we show rigorously in appendix \ref{app:convergence}.

\section{Derivation}
For the derivation it will be useful to introduce
\begin{equation}
\label{tildeG}
	\tilde \cG(x_1, x_2) \colonequals \sqrt{x_{12}}\, \cG(x_1, x_2)
\end{equation}
such that if we `subtract' the discontinuity we obtain
\begin{equation}\label{sDisc}
	\begin{split}
	&\sDisc[\tilde \cG](x_1, x_2) \colonequals \, \tilde \cG(x_1, x_2) - \frac{i}{2} \disc[\tilde \cG](x_2, x_1)\\
	&\,\,\colonequals \, \tilde \cG(x_1, x_2) - \frac{i}{2} \tilde \cG(x_2, x_1 + i \e) + \frac{i}{2} \tilde \cG(x_2, x_1 - i \e)\\
	&\,\,= \, \sqrt{x_{12}} \dDisc[\cG](x_1, x_2)\,,
	\end{split}
\end{equation}
again for $0 < x_2 \leq x_1$. It thus suffices to derive a dispersion relation for $\tilde\cG(x_1, x_2)$ in terms of $\sDisc[\tilde \cG](x_1, x_2)$. Since the $\sDisc[\cdot]$ operation involves the ordinary $\disc[\cdot]$ operation, it is natural to start with a single-variable dispersion relation:
\begin{equation}\label{contact1}
	\tilde \cG(x_1, x_2) = \int_{x_1^-}^\infty \frac{d x_3}{2 \pi i} \frac{\disc[\tilde \cG](x_1, x_3)}{x_{32}}\,,
\end{equation}
valid for $x_2 < x_1$, and the notation $x_1^-$ for the lower bound means $\lim_{\e \searrow 0} \int_{x_1 - \epsilon}(\cdots)$. In appendix \ref{subapp:singlevar} we explain that large $|x_3|$ is nothing but the Regge limit, so this dispersion relation holds for any Regge-bounded correlator.

We now rewrite equation \eqref{contact1} trivially as
\begin{equation}\label{contact2}
	\tilde \cG(x_1, x_2) = \int_{x_1^-}^\infty \frac{d x_3}{\pi x_{32}}  \Big[ \sDisc[\tilde \cG](x_3, x_1) - \tilde \cG(x_3, x_1)\Big]\, .
\end{equation}
Before we can split the right-hand side into two integrals we introduce a regulator to avoid integrating $\tilde \cG(x_3, x_1)$ across its lightcone singularity at $x_3 = x_1$. We introduce, for some small $\eta > 0$, the notation
\begin{equation}
	x_i^\eta \colonequals (1- \eta) x_i\,,	\qquad i = 1,2,3,4\,,
\end{equation}
and then define the $\eta$-regulated subtracted discontinuity
\begin{multline}\label{dDisceta}
		\sDisc^{(\eta)}[\tilde \cG](x_1, x_2) \colonequals \\
		\tilde \cG(x_1, x_2^\eta) - \frac{i}{2} \tilde \cG(x_2, x_1^\eta + i \e) + \frac{i}{2} \tilde \cG(x_2, x_1^\eta - i \e)\,.
\end{multline}
Recall that we will only integrate over $0 < x_2 \leq x_1$. In this region the first term is strictly analytic in $x_2$, since the $\eta$ regulator moves the branch cut from $x_2 \geq x_1$ to $x_2 \geq x_1 / (1- \eta) > x_1$. The other two terms do cover the branch point, but this is not a problem since we interpret them as tempered distributions in $x_1$ after sending $\epsilon \searrow 0$. The limit $\eta \searrow 0$ then has to be taken last. We discuss the $\eta$ regulator in more detail in appendix \ref{app:convergence}.\footnote{The regulator also automatically fixes the well-known subtlety with contact terms in the $\dDisc[\cdot]$ for integer scaling dimensions.}

With the regulator in place we can safely replace equation \eqref{contact2} with\footnote{The left-hand side is actually $\tilde \cG(x_1, x_2^\eta)$, but we will ignore this difference as it is $O(\eta)$.}
\begin{multline}\label{contact3}
	\tilde \cG(x_1, x_2) = \int_{x_1}^\infty \frac{d x_3}{\pi x_{32}}  \sDisc^{(\eta)}[\tilde \cG](x_3, x_1) \\
	- 
	\int_{x_1}^\infty \frac{d x_3}{\pi x_{32}} \tilde \cG(x_3, x_1^\eta)\, .
\end{multline}
In order to arrive at a dispersion relation involving just $\sDisc^{(\eta)}[\tilde \cG]$ we need to cancel the last term. We therefore must introduce another kernel $K_B$ such that:
\begin{align}\label{contactandbulk}
	&- \int_{x_1}^\infty \frac{d x_3}{\pi x_{32}} \tilde \cG(x_3, x_1^\eta) 
	= \\ &\int_{\alpha}^\infty d x_4 \int^{\infty}_{x_4} dx_3 \, K_B(x_1, x_2, x_3, x_4) \sDisc^{(\eta)}[\tilde \cG](x_3, x_4)\nn
\end{align}
for some lower bound $\alpha > 0$ that we will determine imminently. We can unpack the $\sDisc^{(\eta)}[\tilde \cG]$ and change variables to rewrite the right-hand side as 
\begin{align}\label{bulkrhs}
	&\int_{\alpha}^\infty\!\! d x_3 \int^{x_3}_\alpha \!\!dx_4 \, K_B(x_1, x_2, x_3, x_4) \tilde \cG(x_3, x_4^\eta) \\ &- \frac{i}{2} \int^{\infty}_\alpha \!\!dx_3 \int_{x_3}^\infty\!\! d x_4  \, K_B(x_1, x_2, x_4, x_3) \disc[\tilde \cG](x_3, x_4^\eta)\,.\nn
\end{align}
If the resulting equation is to hold for any $\tilde \cG(x_3, x_4)$ which is not necessarily smooth in $x_3$ then it must hold pointwise in $x_3$. We therefore set $\alpha = x_1$ and remove the $x_3$ integral to arrive at:
\begin{multline}\label{bulk2ndeq}
	 -\frac{\tilde \cG(x_3, x_1^\eta)}{\pi x_{32}}  
	 =  \int^{x_3}_{x_1} \!\!dx_4 \, K_B(x_1, x_2, x_3, x_4) \tilde \cG(x_3, x_4^\eta) \\  - \frac{i}{2}  \int_{x_3}^\infty \!\!d x_4  \, K_B(x_1, x_2, x_4, x_3) \disc[\tilde \cG](x_3, x_4^\eta) \, .
\end{multline}
The final integral leads us to postulate analyticity properties of $K_B(x_1, x_2, x_4, x_3)$ in its third argument, so $x_4$. Indeed, it is natural for this function to have a distributional discontinuity from $x_1$ to $x_3$ and to be bounded by a multiple of $|x_4|^{-1}$ at infinity, since this would allow us to integrate the $\disc[\cdot]$ operation by parts with a contour deformation argument. Supposing this, we can remove the $x_4$ integrals also and arrive at the condition
\begin{multline}
	\label{KBeqn}
	- \frac{i}{2} \disc_{x_4} [K_B](x_1, x_2, x_4, x_3) = \\ \frac{\delta(x_{41})}{\pi x_{32}} + K_B(x_1, x_2, x_3, x_4)
\end{multline}
valid as a distribution for $x_4$ in the closed interval $[x_1, x_3]$. Below we will solve equation \eqref{KBeqn} to find $K_B$. In appendix \ref{app:convergence} we will prove that all the above steps are valid manipulations between absolutely convergent integrals, thanks to Regge-boundedness and the $\eta$ regulator.

\subsection{Symmetries}
Equation \eqref{KBeqn} is an equation for a function which is analytic in the $x_4$ plane outside a cut at $[x_1, x_3]$. Together with $x_2$ this gives three marked points in the $x_4$ plane, precisely what we would have for a conformal four-point function in one dimension. (The order is always $x_2 < x_1 < x_3$.) Furthermore, the inhomogeneous term in equation \eqref{KBeqn} transforms covariantly under conformal transformations. As $x_i \mapsto \frac{a x_i + b}{c x_i + d}$ with $ad - bc = 1$, we find that
\begin{equation}
	\frac{\delta(x_{41})}{\pi x_{32}} \to \p{\, \prod_{i = 1}^4 (c x_i + d)} \frac{\delta(x_{41})}{\pi x_{32}} \,.
\end{equation}
It is therefore natural to assume that
\begin{multline}
	K_B\Bigg(\frac{ax_1 + b}{c x_1 + d}, \frac{ax_2 + b}{c x_2 + d}, \frac{ax_3 + b}{c x_3 + d}, \frac{ax_4 + b}{c x_4 + d}\Bigg) \\ = \p{\, \prod_{i = 1}^4 (c x_i + d)} K_B(x_1, x_2, x_3, x_4) 
\end{multline}
as well. As every child knows, this implies that
\begin{equation}\label{KBfoundinx}
	K_B(x_1, x_2, x_3, x_4) =x_{12}^{-1} \, x_{34}^{-1} \, F(x)
\end{equation}
where
\begin{equation}\label{xdef}
	x = \frac{x_{12} x_{34}}{x_{13} x_{24}}
\end{equation}
is the `magic variable' of \cite{Carmi:2019cub}, now seen to be a simple cross-ratio. The points where $x_4$ is equal to $x_2$, $x_1$ and $x_3$ are mapped respectively to $\infty$, $1$ and $0$ in the $x$ variable.

The aforementioned properties of $K_B(x_1, x_2, x_3, x_4)$ translate into analyticity of $F(x)$ away from a cut along the negative real axis, as well as boundedness at infinity. For the discontinuity across this cut equation \eqref{KBeqn} stipulates that, for $x \in [0,1]$,
\begin{equation}
	\label{Fxdisc}
	- \frac{i}{2} \disc_{x} [F]\left(\frac{x}{x-1}\right) = \frac{\delta(x - 1)}{\pi} + F(x)\, .
\end{equation}
Since $F(x)$ is analytic near $x =1$, the Dirac delta must come from a simple pole on the left-hand side, using $\disc[(x-1)^{-1}] = - 2 \pi i \delta(x - 1)$. We thus require that
\begin{equation}
	\label{Fasympt}
	F(x)  \overset{|x| \to \infty}{\sim} - x/\pi^2\,.
\end{equation}

\subsection{Determining \texorpdfstring{$F(x)$}{F(x)}}
We recall that the conformal blocks
\begin{equation}
	k^{(a,b)}_h(x) \colonequals x^{h+a} {}_2 F_1 (h +a, h+b, 2h, x)
\end{equation}
transform nicely under $x \to \frac{x}{x-1}$, to wit
\begin{equation}
	\label{disck}
	k^{(a,b)}_h\p{x/(x-1) \pm i \e}  = e^{\pm i \pi (h+a)} k^{(a,-b)}_h(x)\,.
\end{equation}
We now set $b = 0$ to diagonalize this action. At large $|x|$ we find that
\begin{equation}
	k^{(a,0)}_h(x) \overset{|x| \to \infty}{\sim} \# x^{a} + \# + \ldots\,,
\end{equation}
so we will set $a = 1$ to match equation \eqref{Fasympt}. Now recall that the conformal blocks have a (logarithmic) branch cut at $x > 1$ but $F(x)$ needs to be analytic there. To cancel this cut, to all orders in $(1-x)$, we must take a linear combination with the shadow block, so we define
\begin{align}
	\Psi_{h}^{(1,0)}(x) \colonequals \# k^{(1,0)}_h(x) + \# k^{(1,0)}_{1-h}(x)\,,
\end{align}
with coefficients chosen such that we can also write
\begin{equation}
	\Psi_{h}^{(1,0)}(x) = x^{h + 1} {}_2 F_1(h, h +1, 2, 1-x)\,,
\end{equation}
which is manifestly analytic around $x = 1$. According to \cite{Hogervorst:2017sfd} the family with $h = 1/2 + i \R$ forms a delta-function normalizable basis for an $L^2$ function space on $x \in (0,1)$, so we are not losing any generality at this stage.

Finally we can discuss the discontinuity, equation \eqref{Fxdisc}. Using \eqref{disck} for $x \in (0,1)$ this leads to the condition
\begin{align}
	h = \frac{1}{2} + 2 \N\,.
\end{align}
As we discuss in appendix \ref{subapp:ibpofdisc}, for $h > 1/2$ the behavior at $x = 0$ is too singular for the integrals to make sense. Keeping track of the normalization, we thus arrive at
\begin{equation}\label{Fxfound}
F(x) = - (4\pi)^{-1}\Psi_{\frac{1}{2}}^{(1,0)}(x) 
\end{equation}
as the \emph{unique} solution to the above constraints.

Introducing the $SL(2, \mathbb{R})$ invariant measure
\begin{equation}
	d^2 \mu(x_3, x_4) \colonequals \frac{x_{12}}{x_{34} \, x_{13} \, x_{24}} d x_3 \, d x_4\, ,
\end{equation}
we have thus obtained that the correlator $\tilde \cG$ can be uniquely expressed as
\begin{widetext}
\begin{equation}
	\label{final}
	x_{12} \, \tilde \cG(x_1, x_2) = \iint_{x_1 \leq x_4 \leq x_3} d^2 \mu(x_3, x_4) \, \left(\frac{\delta(x-1)}{\pi}   - \frac{1}{4 \pi} \, x^{\frac12} \, _2 F_1 \left(\frac12,\frac32,2,1-x\right) \right) x_{34}\sDisc^{(\eta)}[\tilde \cG](x_3, x_4)\, .
\end{equation}
\end{widetext}
The power $x_{34}$ can be moved inside the $\sDisc^{(\eta)}[\cdot]$ operator, so the kernel for the function $x_{12} \tilde \cG(x_1, x_2)$ is fully $SL(2, \mathbb{R})$ invariant. In appendix \ref{app:backtostandard} we discuss how this equation is equivalent to the original CDR found in \cite{Carmi:2019cub}.

\section{Conclusion}
We have demystified the CDR by showing that it can be derived using elementary single-variable complex analysis. The contact term in the kernel arises from a single-variable dispersion relation. The bulk term, on the other hand, exhibits a hidden conformal symmetry: it is essentially just a single shadow-symmetric conformal block, with the magical variable $x$ arising as the associated cross-ratio. In the process we have discovered that $\cG(x_1, x_2)$ needs to be analytic only in $x_2$. Our derivation thereby provides an answer to all the questions raised in the introduction.

It would be interesting to apply our methods to the kernels for correlation functions of unequal external dimensions analyzed recently in \cite{Carmi:2025mtx}, or even to find dispersion relations for higher-point functions. It would also be desirable to develop a spacetime picture, as was done in \cite{Simmons-Duffin:2017nub} for the LIF, where $x$ emerges directly as a cross-ratio.
	
\section*{Acknowledgments}
We would like to thank Chris Beem, Dean Carmi, Simon Caron-Huot, Jeremy Mann, and Dalimil Maz\'a\v{c} for discussions. We thank Francesco Russo for collaboration at the initial stages of this project. This work was performed in part at the Aspen Center for Physics, which is supported by National Science Foundation grant PHY-2210452. We acknowledge funding from the European Union (ERC “QFTinAdS”, project number 101087025).

\appendix

\section{Back to the standard dispersion relation}
\label{app:backtostandard}
In the main text we derived the closed-form expression \eqref{final} for $\tilde\cG$, assuming crossing symmetry and in unfamiliar variables. In this appendix we show that it is equivalent to the CDR in terms of the $u$, $v$ cross-ratios of \cite{Caron-Huot:2020adz}.

Equations \eqref{final} and \eqref{sDisc} together imply that the kernel $\cK$ of equation \eqref{originalCDR} for a crossing symmetric $\cG$ reads
\begin{multline}
		\label{fullkernel}
\cK(x_1,x_2,x_3,x_4) = \frac{x_{31}^{1/2}}{\pi\, x_{32} \, x_{12}^{1/2}} \delta(x_{41})\\ - \frac{1 }{4 \pi \, x_{12}^{3/2} \, x_{34}^{1/2}}  x^{\frac32} \, _2 F_1 \left(\frac12,\frac32,2,1-x\right) \theta(x_{41})\, .
\end{multline}
In terms of variables $y_i$, defined as
\begin{equation}
	y_i \colonequals \sqrt{x_i}, \qquad i = \{1,2,3,4\}\,,
\end{equation}
equation \eqref{originalCDR} becomes
\begin{multline}
	\cG(y_1, y_2) = \int_{y_1}^\infty d y_3 \int_{y_1}^{y_3} dy_4 	\; (4 y_3 y_4)\, \cK(y_1,y_2,y_3,y_4)\\	\times \dDisc^{(\eta)}_s[\cG](y_3, y_4)\, ,
\end{multline}
where the factor in front is simply the Jacobian. An analogous relation holds for the antisymmetric correlator, obtained directly from \eqref{antisymmcorr}. Our goal is to combine the symmetric and antisymmetric pieces into a single dispersion relation for a full correlator $\hat\cG$. To do so, note that under $y_4 \to -y_4$ the $s$-channel double discontinuity maps to $\pm\dDisc_t$, with the sign depending on whether the correlator on which the $\dDisc_s$ acts is symmetric or antisymmetric. Introducing
\begin{equation}
	\begin{split}
		K(y_1,y_2,y_3,y_4) \colonequals 2 \left( y_3 y_4 + y_1 y_2 \right) \,  \cK(y_1,y_2,y_3,y_4)  \, ,
	\end{split}
\end{equation}
the full correlator $\hat\cG$ can be reconstructed from both its $s$- and $t$-channel double discontinuities as
\begin{multline}
	\label{fullcorr}
		\hat \cG(y_1, y_2) = \\
		\int_{y_1}^\infty\! d y_3 \int_{y_1}^{y_3}\! dy_4\, \Big(
	 K(y_1,y_2,y_3,y_4) \dDisc_s^{(\eta)}[\hat \cG](y_3, y_4)
	 \\+   K(y_1,-y_2,y_3,y_4) \dDisc_t^{(\eta)}[\hat \cG](y_3, -y_4)\Big) \, .
\end{multline}
We now relabel $y_4 \to -y_4$ in the second term and translate back to the
variables $u$ and $v$. The kernel simply picks up a Jacobian factor,
\begin{equation}
	\hat\cK(u,v,u',v') \colonequals \frac{1}{2 \sqrt{u'} \sqrt{v'}}\, K(u,v,u',v')\, ,
\end{equation}
while the theta function $\theta(y_4 - y_1)$ becomes $\theta(\sqrt{v'} - \sqrt{u'} - \sqrt{u} - \sqrt{v})$. We thus get
\begin{multline}
\hat \cG(u, v) =\\  \int_{u',v'>0} d u' \,  d v' \,  \hat\cK(u,v,u',v')\dDisc_s^{(\eta)}[\hat \cG](u', v')  \\ + \int_{u',v'>0}  du'\, d v' \,  \hat\cK(v,u,v',u') \dDisc_t^{(\eta)}[\hat \cG](u', v') \, ,
\end{multline}
 which as promised coincides with the standard $u$-channel dispersion relation of \cite{Caron-Huot:2020adz}.\footnote{The $k$-subtracted kernels of equation (4.88) of \cite{Caron-Huot:2020adz} can be found by running our argument for the rescaled correlator $x_{12}^{-k/2}\,\cG(x_1,x_2)$, \emph{mutatis mutandis}.} 

\section{Convergence}
\label{app:convergence}
In this section we justify every manipulation in the main text for physical correlation functions respecting the CFT unitarity constraints. In most cases we do so by uniformly bounding the integrands and then appealing to the dominated convergence theorem.

As we mentioned in the introduction, for a general conformal four-point function of identical scalar operators we need to introduce a subtraction as in equation \eqref{subtractionmaintext}. Combining this with equation \eqref{tildeG} this means that
\begin{equation}\label{subtraction1}
	\tilde \cG(x_3, x_4) = x_{34}^{-1/2} \cG^\text{phys}(x_3, x_4)
	\,
\end{equation}
where $\tilde \cG(x_3, x_4)$ is the function that appears in all the integrands in the main text. 

We comment that the need for subtractions is well-known in the literature; it is directly related to the fact that the LIF does not generally converge down to spin 0 \cite{Caron-Huot:2017vep}. We also note that our crossing symmetric subtraction is slightly different from those suggested in \cite{Carmi:2019cub,Caron-Huot:2020adz}.

As can be gleaned from the integration domain sketched in figure \ref{fig:regionsab}, it will be essential for us to bound $\cG^\text{phys}(x_3,x_4)$ at large $x_3$ and/or $x_4$, so at large $v$ and/or $|u|$. Since we normally describe lightcone and OPE limits by sending $u$ and/or $v$ to $0$, we will in this appendix apply a crossing move $(u,v) \mapsto (v/u, 1/u)$ and replace equation \eqref{xfromuv} with
\begin{equation}\label{xaftercross}
	x_3 = \frac{(1 + \sqrt{v})^2}{u}, \qquad x_4 =  \frac{(1 - \sqrt{v})^2}{u},
\end{equation}
In this way it is for example easy to see that the lightcone limit, which is $u \to 0$ at fixed $v$, is simply large $x_3$ and $x_4$ at fixed $x_3/ x_4$.\footnote{After this crossing move the variable $x_3$ is easily verified to only depend on the product $\rho \bar \rho$ in the radial coordinates of \cite{Hogervorst:2013sma}. Therefore, the complex $x_4$ plane at fixed $x_3$ is also the complex $\eta$ plane used in \cite{Caron-Huot:2017vep,Carmi:2019cub}. It is also the complex $t$ plane at fixed $s$ in the conformal Mandelstam variables of \cite{Komatsu:2020sag}. The integration in the latter variables is sketched in \cite{vanRees:2023fcf}, which also shows how the conformal dispersion relation becomes an ordinary fixed $s$ dispersion relation in the flat-space limit.}

\subsection{The single-variable dispersion relation and the Regge limit}
\label{subapp:singlevar}
The first step in the analysis is the derivation of the single-variable dispersion relation in equation \eqref{contact1}. Recall that this is an integral along the orange line in figure \ref{fig:regionsab}. The arc at infinity can be thrown away if
\begin{equation}
	\label{largex2fixedx1}
	\lim_{|x_2| \to \infty} |\tilde \cG(x_1, x_2)| = 0
\end{equation}
pointwise in $x_1$ and uniformly in the complex $x_2$ plane. In the variables of equation \eqref{xaftercross} this corresponds to $u \to 0$ and $v \to 1$ after rotating $v$ once around 0. After writing $u = z \zb$ and $v = (1- z) (1-\zb)$ we see that this is nothing but the well-known Regge limit where $z, \zb \to 0$ but one of them, say $\zb$, is rotated first around 1. If we introduce $\tilde \zb$ for this second-sheet variable, so $\sqrt{1-\zb} = - \sqrt{1 - \tilde \zb}$, then a point corresponding to $x_1 > 1$ and $x_2 = R e^{i \phi}$ at large $R$ corresponds to
\begin{equation}\label{subszR}
	\begin{pmatrix}z\\ \tilde \zb\end{pmatrix} =\frac{2 e^{-i \phi/2}}{\sqrt{R}}\begin{pmatrix} s_+(x_1) \\ s_-(x_1) \end{pmatrix} + O(R^{-1})
\end{equation}
where $s_\pm(x_1) \colonequals \sqrt{x_1} \pm \sqrt{x_1 - 1}$ are finite, strictly positive for $x_1 > 1$, and obey $s_+(x_1) s_-(x_1) = 1$.\footnote{The coefficients implicit in the big $O$ notation are dependent on $x_1$, which is why our analysis only holds pointwise in that variable.}

As we take $R$ large, always at fixed $x_1$, equation \eqref{subszR} shows that $z$ and $\tilde \zb$ not only become small in modulus but also stay strictly away from the negative real axis. We are thus in the domain where the OPE around $z = \zb = \tilde \zb = 1$ converges, and so we can use a standard OPE argument to bound the second-sheet correlation function in terms of its value on the first sheet. On the first sheet we can then use the OPE around $z = \zb = 0$, which in our conventions is the $u$-channel OPE where the identity operator has coefficient 1. Altogether this yields the $x_1$-dependent bound 
\begin{equation}
	\label{boundGphyslargex2}
	|\cG^\text{phys}(x_1, x_2)| \leq C_{x_1}\,,
\end{equation} 
for any $|x_2| > R_{x_1}$. Keeping in mind our single subtraction as in equation \eqref{subtraction1}, we see that equation \eqref{largex2fixedx1} holds uniformly in $x_2$ and pointwise in $x_1$, with almost an entire $\sqrt{x_{12}}$ to spare.

To fully demonstrate the validity of equation \eqref{contact1} we also need to consider the limit of $\cG^\text{phys}(x_1, x_2)$ as $x_2$ approaches $x_1$ (again at fixed finite $x_1$). This is a simple lightcone limit, corresponding to $v \to 0$ at fixed $u$ in the variables of equation \eqref{xaftercross}. Here the correlation function has a simple power-law behavior $\sim v^{-\Delta}$ due to the unit operator in the corresponding OPE channel, which in our variables implies
\begin{equation}\label{Gphyslightcone}
	\cG^{\text{phys}}(x_1, x_2) \overset{x_2 \to x_1}{\sim} x_{12}^{-\Delta}\,.
\end{equation}
Such a power-law divergence is sufficiently mild to conclude that $\cG^{\text{phys}}(x_1, x_2)$, and similarly also $\tilde \cG(x_1, x_2)$, is a valid distribution in $x_2$. (A nice review of the corresponding mathematical theorem by Vladimirov can be found in \cite{Kravchuk:2020scc}.) In fact, since the entire integrand in equation \eqref{contact1} is locally analytic away from $x_2 = x_1$ we can simply apply a contour deformation and integrate $\tilde \cG(x_1, x_2)$ times the test function over a keyhole contour which encircles the singular point. Either procedure suffices to make the integral in equation \eqref{contact1} well-defined near its lower bound.

We have thus established the validity of equation \eqref{contact1} pointwise in $x_1$.

\subsection{Splitting the first integral and the lightcone limit}\label{sec:splitting1st}
Our next job is to justify the split of equation \eqref{contact3} by demonstrating that each integral converges separately. This is where the $\eta$ regulator becomes important.

To do so we first look at the second term of equation \eqref{contact3}, which we recall is given by
\begin{equation}
	\label{contact34appendix}
	\int_{x_1}^\infty \frac{d x_3}{\pi x_{32}} \tilde \cG(x_3, (1- \eta) x_1)\,,
\end{equation}
and corresponds to an integration along the blue line in figure \ref{fig:regionsab}. As we mentioned in the main text the $\eta$ regulator again ensures convergence near the lower end, since $x_2 \mapsto \tilde \cG(x_1, x_2)$ is analytic in $x_2$ unless $x_2 \geq x_1$, but in equation \eqref{contact34appendix} this condition is never met for any $\eta > 0$.

To demonstrate convergence at infinity we need to constrain the behavior of $\cG^{\text{phys}}(x_1, x_2)$ at large positive $x_1$ and fixed $0 < x_2 < x_1$. In terms of the variables of equation \eqref{xaftercross} this is just $u\to 0$, $v \to 1$, so $z, \zb \to 0$, which we recognize to be a simple OPE limit. In our conventions the correlator is bounded by:
\begin{equation}\label{lightconebound}
	|\cG^{\text{phys}}(x_1, x_2)| \leq C_{x_2}
\end{equation}
for $x_1 \geq R_{x_2}$. Note that this first-sheet bound was already used to derive the second-sheet bound of equation \eqref{boundGphyslargex2}. Adding the extra square root from equation \eqref{subtraction1} and the extra $1/x_{32}$ in the integrand in equation \eqref{contact34appendix} implies absolute convergence near infinity, again with almost an entire square root to spare.

For the first term in equation \eqref{contact3} we already noted that the singularity at $x_3 = x_1 /(1-\eta)$ is a valid distribution and can be integrated against $1/x_{32}$ without issue. It thus remains to show that the large $x_3$ limit causes no trouble. But in this region the $\eta$ regulator is unimportant and $\sDisc[\tilde \cG]$ becomes just a linear combination of $\tilde \cG$ and $\disc[\tilde \cG]$. Since we have already established that both integrands are sufficiently well-behaved at large $x_3$, their sum will not cause any trouble either.

This establishes that all the integrals in equation \eqref{contact3} are well-defined, and the split in passing from equation \eqref{contact1} to equation \eqref{contact3} is justified.

\subsection{Integrating the discontinuity by parts}
\label{subapp:ibpofdisc}
We now skip ahead to equation \eqref{bulk2ndeq}, leaving some subtleties with equation \eqref{contactandbulk} for the next subsection. For convenience, we repeat equation \eqref{bulk2ndeq} here (with a question mark):
\begin{multline}\label{KBappendixeq}
		 -\frac{\tilde \cG(x_3, (1 - \eta) x_1)}{\pi x_{32}}  
	 \overset{?}{=} \\ \int^{x_3}_{x_1} \!\!dx_4 \, K_B(x_1, x_2, x_3, x_4) \tilde \cG(x_3, (1-\eta) x_4) \\  - \frac{i}{2}  \int_{x_3}^\infty \!\!d x_4  \, K_B(x_1, x_2, x_4, x_3) \disc[\tilde \cG](x_3, (1-\eta)x_4) \, .
\end{multline}
We will suppose as always that $x_2 < x_1 \leq x_3$. In the following we first consider $x_1 < x_3$ strictly, and then show that the limit $x_3 \searrow x_1$ causes no issues.

The first integral in equation \eqref{KBappendixeq} ranges over $x_1 \leq x_4 \leq x_3$ at fixed $x_3$, which is a finite northwest-pointing line segment somewhere in the blue region in figure \ref{fig:regionsab}. Its analysis is straightforward. We recall that $K_B(x_1, x_2, x_3, x_4)$ is analytic in $x_4$ except for a branch cut at $x_4 \geq x_3$, which touches the upper end of the integration region, and another branch cut at $x_4 \leq x_2$ which we do not probe because $x_2 < x_1$ strictly. Near the relevant branch point it has asymptotic behavior
\begin{align}
	\label{KBlightconeasymptotics}
	K_B(x_1, x_2, x_3, x_4) &\overset{x_4 \to x_3}{\sim}  C_{x_1, x_2} x_{34}^{1/2} \log(x_{34})\, .
\end{align}
We already discussed that $\tilde \cG(x_3, x_4)$ also has a lightcone singularity and associated power-law divergence at $x_4= x_3$, see equation \eqref{Gphyslightcone}. However, the $\eta$ regulator again ensures that we never actually hit this point and $\tilde \cG(x_3, (1-\eta) x_4)$ is analytic and hence uniformly bounded for $x_1 \leq x_4 \leq x_3$. Since the non-analytic behavior of the kernel in equation \eqref{KBlightconeasymptotics} is absolutely integrable, we may conclude that the first integral in equation \eqref{KBappendixeq} is completely safe.

The second integral corresponds again to a northwest-pointing line in figure \ref{fig:regionsab}, but now it lies in the orange region and is semi-infinite. We recall that we not only wish to establish the convergence of this integral but also to justify the `integration by parts' of the $\disc[\cdot]$ operation from $\tilde \cG(x_3, (1-\eta) x_4)$ to $K_B(x_1, x_2, x_4, x_3)$.

Let us first discuss the large $|x_4|$ behavior. We straightforwardly find the bound
\begin{equation}\label{KBsecondlimit}
	|K_B(x_1,x_2,x_4,x_3) | \le \frac{D_{x_1,x_2,x_3}}{|x_4|} \, ,
\end{equation}
for any $|x_4| > R_{x_1, x_2, x_3}$. The asymptotic behavior of $K_B$ is therefore the same as the simple power in an ordinary single-variable dispersion relation. We already established equation \eqref{boundGphyslargex2} by Regge-boundedness, which in turn implies at least a square root falloff of $\tilde \cG(x_3, x_4)$ at large $|x_4|$. Therefore not only does the second integral in equation \eqref{KBappendixeq} converge absolutely at large $x_4$, but we also find that the contribution from the arc at infinity vanishes if we integrate the $\disc[\cdot]$ by parts, as before with almost an entire square root to spare.

This leaves the behavior at small $x_{34}$ of the second integral in equation \eqref{KBappendixeq}. At this point the $\eta$ regulator is again essential. The kernel is simple enough for equation \eqref{KBlightconeasymptotics} to apply again,
\begin{align}\label{KBsecondlimit2}
	K_B(x_1, x_2, x_4, x_3) &\overset{x_4 \to x_3}{\sim}  C_{x_1, x_2} x_{34}^{1/2} \log(x_{34})\,.
\end{align}
Let us first consider the point $x_4 = x_3$ and then the point $x_4 = x_3/(1-\eta)$.

As before, at $x_4 = x_3$ the function $\tilde \cG(x_3,(1-\eta)x_4)$ is analytic and the integral itself causes no problems since the singular behavior of equation \eqref{KBsecondlimit2} is integrable. Now, however, we draw the further conclusion that we do not pick up any additional contribution if we rotate the contour in the complex $x_4$ plane from $x_4 > x_3$ to $x_4 < x_3$ in the integration by parts of the $\disc[\cdot]$ operation.

At $x_4 = x_3/(1-\eta)$ we find the aforementioned lightcone singularity of $\tilde \cG(x_3, x_4)$, but here we are a finite distance away from the endpoint and so $K_B(x_1, x_2, x_4, x_3)$ is analytic. As before we can either apply a small contour deformation or simply interpret $\tilde \cG(x_3, (1-\eta) x_4)$ as a distribution to claim that the integral is also well-defined around this point.

We have thus established the pointwise validity of equation \eqref{KBappendixeq}, which was the same as equation \eqref{bulk2ndeq} in the main text, and also that the $\disc[\cdot]$ operation can safely be integrated by parts without contributions from large $|x_4|$ or small $|x_{34}|$.

The above derivation highlights the main purpose of the $\eta$ regulator, which is to separate the branch cuts of $K_B$ and $\tilde \cG$. In this way one of them is always analytic even if the other is not, and a distribution in our integrals is always paired against an analytic function.

\subsubsection{More singular kernels}\label{sec:singkernels}
In the main text we briefly discussed the possibility of introducing other kernels given in terms of the $\Psi_{h}^{(1,0)}(x)$ with $h = \frac{1}{2} + 2k$ and $k > 0$. We can for example try 
\begin{equation}\label{familyofsol}
	F^{(k)}(x) = - \frac{1- 16 k^2}{4 \pi}\Psi_{\frac{1}{2} +  2 k}^{(1,0)}(x)\, ,
\end{equation}
or linear combinations thereof, with an overall normalization such that equation \eqref{Fasympt} is obeyed. Note however that the behavior around 0 for $k > 0$ is given by\footnote{For $k = 0$ there is an extra $\log(x)$ in the numerator. For $k > 0$ the same $\log(x)$ appears with a sub-leading power of $x$.} 
\begin{equation}
	\Psi_{\frac{1}{2} + 2k}^{(1,0)}(x) \overset{x\to 0}{\sim} \frac{\#}{x^{2k -3/2}}
\end{equation}
and it is therefore increasingly singular for larger $k$. In fact, for $k > 0$ one has
\begin{align}
	\label{Kbksingular}
		K_B^{(k)}(x_1,x_2,x_3,x_4)  \overset{x_4\to x_3}{\sim} E_{x_1, x_2,x_3}\, x_{34}^{1/2 - 2 k} \, , 
\end{align} 
with $E$ being a positive coefficient depending on $x_1,x_2,x_3$, so that the integrand is no longer integrable around $x_4 = x_3$ for $k \neq 0$, and even if it were, the contour rotation may no longer be justified. This limit thus singles out $k=0$.

Of course one might run into physical correlation functions $\cG^\text{phys}(x_3, x_4)$ with very soft behavior in the lightcone limit $x_{34} \to 0$. In that case one could try to remove the $\eta$ regulator and even introduce more singular kernels with $k > 0$, which is valid provided the correlation function falls off rapidly enough to offset the divergent behavior of equation \eqref{Kbksingular}. We however stress that one should also control the \emph{double} lightcone limit of $\cG^\text{phys}(x_3, x_4)$, as we will see below.

\subsubsection{The contact point limit}\label{sec:contactpoint}
Recall that equation \eqref{bulk2ndeq} is supposed to be integrated over $x_3$ from $x_1$ to $\infty$ as in equation \eqref{bulkrhs}. The limit $x_3 \searrow x_1$ deserves special attention because the two endpoints of the first integral in \eqref{KBappendixeq} collide. Substituting it naively is ambiguous: the integration range shrinks to zero, while the kernel diverges. Both effects are absorbed by the Jacobian if we trade $x_4$ for the cross-ratio $x$, whose range is $[0,1]$ for any $x_3$. Explicitly, the integral becomes
\begin{equation}\label{contactx3tox1}
	\int_0^1 d x \frac{F(x)}{x} \frac{\tilde \cG(x_3, (1-\eta)x_4(x))}{x_{12} + x \, x_{31}}\, .
\end{equation}
It is now completely safe to set $x_3 = x_1$. The variable $x_4(x)$ reduces to $x_1$ for every $x$ and we obtain
\begin{multline}\label{onefouronepi}
\frac{\tilde \cG(x_1, (1-\eta)x_1)}{x_{12}} \int_0^1\!\! d x \frac{F(x)}{x} =\frac{\tilde \cG(x_1, (1-\eta)x_1)}{x_{12}} \\ \times \left( \frac{1}{4} - \frac{1}{\pi} \right)\, 
\end{multline}
for the $x_3 \to x_1$ limit of the first integral.

Let us now turn to the second integral in equation \eqref{KBappendixeq}. For $x_3 = x_1$ the kernel collapses to the Cauchy one. Changing the integration variable from $x_4$ to $(1-\eta)\, x_4$, the integral reduces to 
\begin{align}
	- \frac{1}{4 x_{12}}	\int_{(1-\eta) x_1}^{\infty} \frac{d x_4}{2\pi i} \frac{\disc[\tilde \cG](x_1, x_4)}{x_{4} - (1-\eta)x_1}  \, .
\end{align}
The pole of the kernel now sits at $x_4 = (1-\eta) x_1$, a finite distance $\eta x_1$ below the branch point at which the discontinuity starts. This expression is precisely the
single-variable dispersion relation \eqref{contact1} evaluated at $x_2 = (1-\eta) x_1$, whose convergence at large $x_4$ is guaranteed by the single subtraction. The integral can therefore be safely evaluated to
\begin{equation}
-	\frac{1}{4} \frac{\tilde \cG(x_1, (1-\eta)x_1)}{x_{12}}\, .
\end{equation} 
We see that each term in equation \eqref{KBappendixeq} continues to make sense in the limit $x_3 \searrow x_1$. Furthermore, the $1/4$ cancels precisely against the corresponding factor in equation \eqref{onefouronepi} to provide the correct left-hand side. We therefore also view the above computation as a non-trivial check of the normalization of $F(x)$.

\subsection{Combining the different terms}\label{sec:combining}
In the previous subsection the two $x_4$ integrals of equation \eqref{bulk2ndeq} were shown to converge separately, pointwise in $x_3$. The final steps are to restore the $x_3$ integral as in equation \eqref{bulkrhs} and to justify the swap of the $x_3$ and $x_4$ integrals in the first of these integrals, which is necessary to pack the two terms in equation \eqref{bulkrhs} into a single term involving the $\sDisc^{(\eta)}[\cdot]$.

This part of the analysis is the most delicate. We are now probing the correlation function $\cG^\text{phys}(x_1, x_2)$ across the shaded domain of figure \ref{fig:regionsab}, so in particular we need to understand the region where both $x_1 \to \infty$ and $x_2 \to \infty$, whilst their ratio $x_1 / x_2 \in [0, \infty]$. This includes the notorious double lightcone limit where $x_1/x_2 \to 1$, which is $u, v \to 0$ in the variables of equation \eqref{xaftercross}. We will see that the $\eta$ regulator is essential to make the integrals well-defined.

We first investigate the first integral of \eqref{bulkrhs}, which is
\begin{equation}
	\label{firstintegralbulkrhsapp}
\int_{x_1}^\infty\!\! d x_3 \int^{x_3}_{x_1} \!\!dx_4 \, K_B(x_1, x_2, x_3, x_4) \tilde \cG(x_3, (1-\eta) x_4)\,\,,
\end{equation}
and corresponds to the blue domain of figure \ref{fig:regionsab}. We will show that it is absolutely convergent.

We again pass to the variables of equation \eqref{xaftercross}, which with the $\eta$ regulator yield
\begin{equation}\label{uvaftercrossx3}
	\sqrt{v} =
	\frac{1 - \sqrt{(1-\eta)\,x_4/x_3}}{1 + \sqrt{(1-\eta)\,x_4/x_3}}\,,
	\quad
	u = \frac{(1+\sqrt v)^2}{x_3} \,\leq\, \frac{4}{x_3}\,.
\end{equation}
The delicate region is that of large $x_3$, which corresponds to the lightcone limit $u \to 0$ with $v$ positive and on the first sheet. Importantly, because of the $\eta$ regulator we have the lower bound $\sqrt{v} \geq \#\,\eta$ across the integration domain. In terms of $z$ and $\zb$ this means $0 \leq z \leq \#\, x_3^{-1/2}$ and $0 \leq \bar z \leq 1 - \#\,\eta^2$. We are once more strictly inside the region of convergence of the OPE around $z = \bar z = 0$, and can thus bound the correlator by its value at a corner of this region:
\begin{equation}\label{Cetabound}
	\big|\cG^{\rm phys}\big(x_3, (1-\eta)x_4\big)\big| \,\leq\, C_\eta\,.
\end{equation}
Here $C_\eta$ is a positive constant independent of $x_3$ and $x_4$, and it is finite because $\bar z$ is always strictly smaller than $1$ for $\eta > 0$. Enlarging $C_\eta$ if necessary and using that $x_3 - (1-\eta)x_4 \geq \eta\, x_3$, this directly gives the bound
\begin{equation}\label{GtildeCetabound}
	\big|\tilde\cG\big(x_3,(1-\eta)x_4\big)\big| \,\leq\,
	\frac{C_\eta}{\sqrt{\eta\, x_3}}\,,
\end{equation}
across the entire integration domain of equation \eqref{firstintegralbulkrhsapp}.

Next we bound the kernel. First, we trade $x_4$ for the cross-ratio $x$. As in equation \eqref{contactx3tox1}, the range $x_4 \in [x_1,x_3]$ becomes $x \in [0,1]$ for every $x_3$. The prefactor can be bounded over the whole domain by 
\begin{equation}
	x_{12} + x\,  x_{31} \ge x_{12}^{1/4} (x \, x_{31})^{3/4}\, , 
\end{equation}
while the function $F(x)$, which we remember behaves as $x^{3/2} \log(x)$ at small $x$ and is analytic elsewhere on $[0,1]$, can be bounded by
\begin{equation}\label{boundF}
	|F(x)| \le \frac{1}{\pi} x^{3/2} \left( 1 + |\log(x)| \right)\, .
\end{equation}
Altogether, the integrand is bounded by the product
\begin{equation}
\frac{	C_\eta }{\pi \, \sqrt{\eta} \, x_{12}^{1/4}} \, \,  x_3^{-1/2} x_{31}^{-3/4} \, x^{-1/4} \left( 1 + |\log(x)| \right) \,. 
\end{equation}
Each factor is absolutely integrable over its own domain. The double integral \eqref{firstintegralbulkrhsapp} is therefore absolutely convergent and consequently the order of the $x_3$ and $x_4$ integrations does not matter.

Our final task is to show that the second integral of equation \eqref{bulkrhs}, which is
\begin{equation}
	\label{secondintegralbulkrhsapp}
	\int^{\infty}_{x_1} \!\!dx_3 \int_{x_3}^\infty\!\! d x_4  \, K_B(x_1, x_2, x_4, x_3) \disc[\tilde \cG](x_3, (1-\eta)x_4)\,,
\end{equation}
is also well-defined, despite the integration over the orange domain in figure \ref{fig:regionsab} reaching all the way to the double lightcone limit.

We rewrite the integral as
\begin{multline}
	\label{eq:beforesplit}
	\frac{1}{1-\eta}\int^{\infty}_{x_1} \!\!dx_3 \int_{(1- \eta) x_3}^\infty\!\! d x_4  \, K_B\left(x_1, x_2, \frac{x_4}{1-\eta}, x_3\right) \\ \times \disc[\tilde \cG](x_3,x_4)
\end{multline}
and split the $x_4$ integral at $(1+\eta) x_3$. We start with the upper part:
\begin{multline}\label{integraleta2}
	\frac{1}{1-\eta}\int^{\infty}_{x_1} \!\!dx_3 \int_{(1+ \eta) x_3}^\infty\!\! d x_4  \, K_B\left(x_1, x_2, \frac{x_4}{1-\eta}, x_3\right) \\ \times \disc[\tilde \cG](x_3,x_4)\, .
\end{multline}
Both terms here are relatively straightforward to bound. The second-sheet discontinuity $\disc[\tilde \cG](x_3, x_4)$ is bounded in terms of the first-sheet function $\tilde \cG(x_4, x_3)$, which from equation \eqref{GtildeCetabound} is bounded by $C_\eta / \sqrt{x_4}$ across the integration domain. 

As for the kernel, the argument of $F$ is the cross-ratio
\begin{equation}\label{xsecondintegral}
	x = \frac{x_{12} \, (x_4 - (1-\eta)x_3)}{x_{32}\, (x_4 - (1-\eta)x_1)} \leq \frac{x_{12}}{x_{32}} \left(\frac{1+\eta}{2\eta} \right)\, ,
\end{equation} 
where the inequality holds across the integration domain of equation \eqref{integraleta2}. Since also $x \in (0,1]$ throughout this domain, the bound \eqref{boundF} on $F(x)$ still holds. If we replace it (for simplicity) by the weaker bound
\begin{equation}\label{boundsecondintF}
	|F(x)| \le \frac{2}{\pi} x^{5/4}\, ,
\end{equation}
then we can deduce that
\begin{multline}
	\left| K_B\left(x_1, x_2, \frac{x_4}{1-\eta}, x_3\right)\right| \le \\ \frac{2 (1-\eta)}{\pi x_4} \left(\frac{1+ \eta}{2 \eta}\right)^{5/4} \frac{x_{12}^{1/4}}{x_{32}^{5/4}}
\end{multline}
across the whole domain, using also $x_4/(1-\eta) - x_3 \le x_4 / (1-\eta)$ to bound the prefactor in $K_B$.

The integrand of \eqref{integraleta2} is therefore bounded by $x_{32}^{-5/4} x_4^{-3/2}$ times a positive constant independent of $x_3$ and $x_4$ (but depending on $x_1$, $x_2$ and $\eta$). This function is absolutely integrable, so the integral in \eqref{integraleta2} is absolutely convergent.

It remains to show that we can make sense of the second part of the split of equation \eqref{eq:beforesplit}, which is:
\begin{multline}
	\label{dlcintegral}
	\frac{1}{1-\eta}\int^{\infty}_{x_1} \!\!dx_3 \int^{(1+ \eta) x_3}_{(1-\eta) x_3}\!\! d x_4  \, K_B\left(x_1, x_2, \frac{x_4}{1-\eta}, x_3\right) \\ \times \disc[\tilde \cG](x_3,x_4)\,.
\end{multline}
This includes the region $x_4 = x_3$ where we have the power-law bound \eqref{Gphyslightcone}, so for generic $\Delta$ we must interpret the integral distributionally. Previously we did so pointwise in $x_3$, which was straightforward, but now we need a stronger result because we integrate $x_3$ all the way to infinity. We will therefore resort to contour deformation to regulate the integral, after which we will show that we can uniformly bound the integrand.

Consider then the deformation of the contour into the complex $x_4$ plane to a semicircle of radius $\eta x_3$ around $x_4 = x_3$, with the semicircle sitting in the upper or lower half plane depending on which part of the $\disc[\tilde \cG](x_3, x_4)$ we consider. The kernel $K_B(x_1, x_2, x_4/(1-\eta), x_3)$ is analytic in $x_4$ along this contour, and can be uniformly bounded in both variables as follows.

Parametrizing $x_4 = x_3 (1 + \eta e^{i \theta})$, with $\theta \in (0,\pi)$
on the upper semicircle and $\theta \in (-\pi,0)$ on the lower, we have
\begin{equation}
x_4 - (1-\eta)x_3 = 2 \eta x_3 \cos(\theta/2)\, e^{i \theta/2}\, ,
\end{equation}
while 
\begin{equation}
|x_4 - (1-\eta)x_1| \ge \max\{2 \eta x_3 \cos(\theta/2),\, (1-\eta) x_{31}\}\, .
\end{equation}
The phase of $x_4 - (1-\eta)x_1$ has the same sign as $\theta$ and is at most $|\theta|/2$ in modulus. The argument of $F$ in \eqref{xsecondintegral} thus stays in the right half of the closed unit disk, where $F$ obeys
\begin{equation}
	|F(x)| \le \frac{2^{5/4}}{\pi}\, |x|^{5/4}\, ,
\end{equation}
as follows from the Euler integral representation of the hypergeometric
function together with \eqref{boundsecondintF}.

This, together with the bound
 \begin{multline}
 	|x_4 - (1-\eta) x_1|^{-5/4} \le ((1-\eta) \, x_{31})^{-3/4} \\ \times (2 \, \eta\,  x_3 \cos(\theta/2))^{-1/2}\, ,
 \end{multline} 
implies the upper bound
\begin{multline}
	\left| K_B\left(x_1, x_2, \frac{x_4}{1-\eta}, x_3\right)\right| \le \frac{2 \, x_{12}^{1/4}}{\pi \, } \left( \frac{1-\eta}{\eta} \right)^{1/4}\\ \times x_{32}^{-5/4} x_{31}^{-3/4}  x_3^{-1/4} \cos^{-1/4}(\theta/2)\, .
\end{multline}
Below we will prove that, along the contour,
\begin{equation}
	\label{Gphysfinalbound}
	|\cG^\text{phys}(x_3, x_4)| \leq C_\eta
\end{equation}
and therefore, using that $|x_{34}| = \eta x_3$ along the contour, we have the corresponding bound
\begin{equation}
|\tilde \cG(x_3,x_4)| \le \frac{C_\eta}{\sqrt{ \eta x_3}}\, .
\end{equation}
Combining the above two bounds, the integrand of \eqref{dlcintegral} is now bounded in absolute value by a constant times $\cos^{-1/4}(\theta/2)\, x_3^{1/4} x_{32}^{-5/4} x_{31}^{-3/4}$. The two integrations factorize and both converge. (In particular, $\cos^{-1/4}(\theta/2)$ is integrable at the endpoints $\theta = \pm\pi$ of the semicircles.) Therefore, for $0 < \eta < 1$ our target expression is absolutely integrable along the contour. Since the contour can be continuously deformed without changing its value, equation \eqref{dlcintegral} indeed makes sense, albeit by interpreting the inner integral via a contour deformation.

It remains to fulfill our promise and show that equation \eqref{Gphysfinalbound} holds throughout the semi-infinite cylinder given by $x_3 > 1$ and $|x_4 - x_3| = \eta \, x_3$. In the variables of equation \eqref{xaftercross}, this set is given by
\begin{equation}
	\left| \frac{4 \sqrt{v}}{(1 + \sqrt{v})^2}\right| = \eta\, , \qquad \sqrt{u} = \frac{1 + \sqrt{v}}{\sqrt{x_3}}\,.
\end{equation}
For small $\eta$ we see that $\sqrt{v}$ carves out a small circular arc of radius approximately $\eta/4$ around the origin, and as it does so $\sqrt{u}$ follows a similar arc, centered around $1/\sqrt{x_3}$ and with a radius of approximately $\eta / (4 \sqrt{x_3})$. As $x_3$ becomes large, this arc in the $u$ plane approaches the origin and we approach the double lightcone limit where both $\sqrt{u}$ and $\sqrt{v}$ become very small.

Importantly, however, the arc in the $\sqrt{u}$ plane shrinks to the \emph{positive} real axis, and moreover it does so in such a way that we can bound $|\arg(\sqrt{u})\,| < \epsilon_\eta$ for any $\epsilon_\eta > 0$ by picking $\eta$ small enough. But in the $z$ and $\zb$ variables this means that as $\zb$ traces out a small circle around 1, the $z$ variable approaches zero \emph{from the right}. This means that even our complexified contour never touches the dangerous region of negative real $z$, irrespective of whether $\zb$ is on the first or on the second sheet.

It is thus our good fortune that, for sufficiently small $\eta$, we remain in the domain of convergence of the OPE around $z = \zb = 1$. This allows us to use this OPE to bound the correlation function in terms of some values $z_*$ and $\zb_*$ on the real axis. Their exact values can be found using the radial coordinates of \cite{Hogervorst:2013sma}, but we are sure that $0 \leq z_* \leq \zb_* \leq 1$. At the same time, along the contour we remain a finite $\eta$-dependent distance away from $\zb = 1$. We are therefore \emph{strictly} in the domain of convergence of the OPE around $z = \zb = 0$, allowing us to indeed bound the physical correlation function by a constant.

This establishes that both integrals in equation \eqref{bulkrhs}, which we reproduced in this appendix in \eqref{firstintegralbulkrhsapp} and \eqref{secondintegralbulkrhsapp}, are well-defined. This completes our justification of the manipulations in the main text.

\bibliography{biblio}

\begin{thebibliography}{21}%
\makeatletter
\providecommand \@ifxundefined [1]{%
 \@ifx{#1\undefined}
}%
\providecommand \@ifnum [1]{%
 \ifnum #1\expandafter \@firstoftwo
 \else \expandafter \@secondoftwo
 \fi
}%
\providecommand \@ifx [1]{%
 \ifx #1\expandafter \@firstoftwo
 \else \expandafter \@secondoftwo
 \fi
}%
\providecommand \natexlab [1]{#1}%
\providecommand \enquote  [1]{``#1''}%
\providecommand \bibnamefont  [1]{#1}%
\providecommand \bibfnamefont [1]{#1}%
\providecommand \citenamefont [1]{#1}%
\providecommand \href@noop [0]{\@secondoftwo}%
\providecommand \href [0]{\begingroup \@sanitize@url \@href}%
\providecommand \@href[1]{\@@startlink{#1}\@@href}%
\providecommand \@@href[1]{\endgroup#1\@@endlink}%
\providecommand \@sanitize@url [0]{\catcode `\\12\catcode `\$12\catcode
  `\&12\catcode `\#12\catcode `\^12\catcode `\_12\catcode `\%12\relax}%
\providecommand \@@startlink[1]{}%
\providecommand \@@endlink[0]{}%
\providecommand \url  [0]{\begingroup\@sanitize@url \@url }%
\providecommand \@url [1]{\endgroup\@href {#1}{\urlprefix }}%
\providecommand \urlprefix  [0]{URL }%
\providecommand \Eprint [0]{\href }%
\providecommand \doibase [0]{http://dx.doi.org/}%
\providecommand \selectlanguage [0]{\@gobble}%
\providecommand \bibinfo  [0]{\@secondoftwo}%
\providecommand \bibfield  [0]{\@secondoftwo}%
\providecommand \translation [1]{[#1]}%
\providecommand \BibitemOpen [0]{}%
\providecommand \bibitemStop [0]{}%
\providecommand \bibitemNoStop [0]{.\EOS\space}%
\providecommand \EOS [0]{\spacefactor3000\relax}%
\providecommand \BibitemShut  [1]{\csname bibitem#1\endcsname}%
\let\auto@bib@innerbib\@empty
\bibitem [{\citenamefont {Carmi}\ and\ \citenamefont
  {Caron-Huot}(2020)}]{Carmi:2019cub}%
  \BibitemOpen
  \bibfield  {author} {\bibinfo {author} {\bibfnamefont {D.}~\bibnamefont
  {Carmi}}\ and\ \bibinfo {author} {\bibfnamefont {S.}~\bibnamefont
  {Caron-Huot}},\ }\href {\doibase 10.1007/JHEP09(2020)009} {\bibfield
  {journal} {\bibinfo  {journal} {JHEP}\ }\textbf {\bibinfo {volume} {09}},\
  \bibinfo {pages} {009} (\bibinfo {year} {2020})},\ \Eprint
  {http://arxiv.org/abs/1910.12123} {arXiv:1910.12123 [hep-th]} \BibitemShut
  {NoStop}%
\bibitem [{\citenamefont {Caron-Huot}(2017)}]{Caron-Huot:2017vep}%
  \BibitemOpen
  \bibfield  {author} {\bibinfo {author} {\bibfnamefont {S.}~\bibnamefont
  {Caron-Huot}},\ }\href {\doibase 10.1007/JHEP09(2017)078} {\bibfield
  {journal} {\bibinfo  {journal} {JHEP}\ }\textbf {\bibinfo {volume} {09}},\
  \bibinfo {pages} {078} (\bibinfo {year} {2017})},\ \Eprint
  {http://arxiv.org/abs/1703.00278} {arXiv:1703.00278 [hep-th]} \BibitemShut
  {NoStop}%
\bibitem [{\citenamefont {Dobrev}\ \emph {et~al.}(1977)\citenamefont {Dobrev},
  \citenamefont {Mack}, \citenamefont {Petkova}, \citenamefont {Petrova},\ and\
  \citenamefont {Todorov}}]{Dobrev:1977qv}%
  \BibitemOpen
  \bibfield  {author} {\bibinfo {author} {\bibfnamefont {V.~K.}\ \bibnamefont
  {Dobrev}}, \bibinfo {author} {\bibfnamefont {G.}~\bibnamefont {Mack}},
  \bibinfo {author} {\bibfnamefont {V.~B.}\ \bibnamefont {Petkova}}, \bibinfo
  {author} {\bibfnamefont {S.~G.}\ \bibnamefont {Petrova}}, \ and\ \bibinfo
  {author} {\bibfnamefont {I.~T.}\ \bibnamefont {Todorov}},\ }\href {\doibase
  10.1007/BFb0009678} {\emph {\bibinfo {title} {{Harmonic Analysis on the
  n-Dimensional Lorentz Group and Its Application to Conformal Quantum Field
  Theory}}}},\ Vol.~\bibinfo {volume} {63}\ (\bibinfo {year}
  {1977})\BibitemShut {NoStop}%
\bibitem [{\citenamefont {Caron-Huot}\ \emph
  {et~al.}(2021{\natexlab{a}})\citenamefont {Caron-Huot}, \citenamefont
  {Mazac}, \citenamefont {Rastelli},\ and\ \citenamefont
  {Simmons-Duffin}}]{Caron-Huot:2020adz}%
  \BibitemOpen
  \bibfield  {author} {\bibinfo {author} {\bibfnamefont {S.}~\bibnamefont
  {Caron-Huot}}, \bibinfo {author} {\bibfnamefont {D.}~\bibnamefont {Mazac}},
  \bibinfo {author} {\bibfnamefont {L.}~\bibnamefont {Rastelli}}, \ and\
  \bibinfo {author} {\bibfnamefont {D.}~\bibnamefont {Simmons-Duffin}},\ }\href
  {\doibase 10.1007/JHEP05(2021)243} {\bibfield  {journal} {\bibinfo  {journal}
  {JHEP}\ }\textbf {\bibinfo {volume} {05}},\ \bibinfo {pages} {243} (\bibinfo
  {year} {2021}{\natexlab{a}})},\ \Eprint {http://arxiv.org/abs/2008.04931}
  {arXiv:2008.04931 [hep-th]} \BibitemShut {NoStop}%
\bibitem [{\citenamefont {Caron-Huot}\ \emph
  {et~al.}(2021{\natexlab{b}})\citenamefont {Caron-Huot}, \citenamefont
  {Mazac}, \citenamefont {Rastelli},\ and\ \citenamefont
  {Simmons-Duffin}}]{Caron-Huot:2021enk}%
  \BibitemOpen
  \bibfield  {author} {\bibinfo {author} {\bibfnamefont {S.}~\bibnamefont
  {Caron-Huot}}, \bibinfo {author} {\bibfnamefont {D.}~\bibnamefont {Mazac}},
  \bibinfo {author} {\bibfnamefont {L.}~\bibnamefont {Rastelli}}, \ and\
  \bibinfo {author} {\bibfnamefont {D.}~\bibnamefont {Simmons-Duffin}},\ }\href
  {\doibase 10.1007/JHEP11(2021)164} {\bibfield  {journal} {\bibinfo  {journal}
  {JHEP}\ }\textbf {\bibinfo {volume} {11}},\ \bibinfo {pages} {164} (\bibinfo
  {year} {2021}{\natexlab{b}})},\ \Eprint {http://arxiv.org/abs/2106.10274}
  {arXiv:2106.10274 [hep-th]} \BibitemShut {NoStop}%
\bibitem [{\citenamefont {Mack}(2009)}]{Mack:2009mi}%
  \BibitemOpen
  \bibfield  {author} {\bibinfo {author} {\bibfnamefont {G.}~\bibnamefont
  {Mack}},\ }\href@noop {} {\  (\bibinfo {year} {2009})},\ \Eprint
  {http://arxiv.org/abs/0907.2407} {arXiv:0907.2407 [hep-th]} \BibitemShut
  {NoStop}%
\bibitem [{\citenamefont {Penedones}\ \emph {et~al.}(2020)\citenamefont
  {Penedones}, \citenamefont {Silva},\ and\ \citenamefont
  {Zhiboedov}}]{Penedones:2019tng}%
  \BibitemOpen
  \bibfield  {author} {\bibinfo {author} {\bibfnamefont {J.}~\bibnamefont
  {Penedones}}, \bibinfo {author} {\bibfnamefont {J.~A.}\ \bibnamefont
  {Silva}}, \ and\ \bibinfo {author} {\bibfnamefont {A.}~\bibnamefont
  {Zhiboedov}},\ }\href {\doibase 10.1007/JHEP08(2020)031} {\bibfield
  {journal} {\bibinfo  {journal} {JHEP}\ }\textbf {\bibinfo {volume} {08}},\
  \bibinfo {pages} {031} (\bibinfo {year} {2020})},\ \Eprint
  {http://arxiv.org/abs/1912.11100} {arXiv:1912.11100 [hep-th]} \BibitemShut
  {NoStop}%
\bibitem [{\citenamefont {Maz{\'a}{\v{c}}}\ \emph {et~al.}(2021)\citenamefont
  {Maz{\'a}{\v{c}}}, \citenamefont {Rastelli},\ and\ \citenamefont
  {Zhou}}]{Mazac:2019shk}%
  \BibitemOpen
  \bibfield  {author} {\bibinfo {author} {\bibfnamefont {D.}~\bibnamefont
  {Maz{\'a}{\v{c}}}}, \bibinfo {author} {\bibfnamefont {L.}~\bibnamefont
  {Rastelli}}, \ and\ \bibinfo {author} {\bibfnamefont {X.}~\bibnamefont
  {Zhou}},\ }\href {\doibase 10.1007/JHEP08(2021)140} {\bibfield  {journal}
  {\bibinfo  {journal} {JHEP}\ }\textbf {\bibinfo {volume} {08}},\ \bibinfo
  {pages} {140} (\bibinfo {year} {2021})},\ \Eprint
  {http://arxiv.org/abs/1910.12855} {arXiv:1910.12855 [hep-th]} \BibitemShut
  {NoStop}%
\bibitem [{\citenamefont {Sleight}\ and\ \citenamefont
  {Taronna}(2020)}]{Sleight:2019ive}%
  \BibitemOpen
  \bibfield  {author} {\bibinfo {author} {\bibfnamefont {C.}~\bibnamefont
  {Sleight}}\ and\ \bibinfo {author} {\bibfnamefont {M.}~\bibnamefont
  {Taronna}},\ }\href {\doibase 10.1007/JHEP11(2020)075} {\bibfield  {journal}
  {\bibinfo  {journal} {JHEP}\ }\textbf {\bibinfo {volume} {11}},\ \bibinfo
  {pages} {075} (\bibinfo {year} {2020})},\ \Eprint
  {http://arxiv.org/abs/1912.07998} {arXiv:1912.07998 [hep-th]} \BibitemShut
  {NoStop}%
\bibitem [{\citenamefont {Gopakumar}\ \emph {et~al.}(2017)\citenamefont
  {Gopakumar}, \citenamefont {Kaviraj}, \citenamefont {Sen},\ and\
  \citenamefont {Sinha}}]{Gopakumar:2016wkt}%
  \BibitemOpen
  \bibfield  {author} {\bibinfo {author} {\bibfnamefont {R.}~\bibnamefont
  {Gopakumar}}, \bibinfo {author} {\bibfnamefont {A.}~\bibnamefont {Kaviraj}},
  \bibinfo {author} {\bibfnamefont {K.}~\bibnamefont {Sen}}, \ and\ \bibinfo
  {author} {\bibfnamefont {A.}~\bibnamefont {Sinha}},\ }\href {\doibase
  10.1103/PhysRevLett.118.081601} {\bibfield  {journal} {\bibinfo  {journal}
  {Phys. Rev. Lett.}\ }\textbf {\bibinfo {volume} {118}},\ \bibinfo {pages}
  {081601} (\bibinfo {year} {2017})},\ \Eprint
  {http://arxiv.org/abs/1609.00572} {arXiv:1609.00572 [hep-th]} \BibitemShut
  {NoStop}%
\bibitem [{\citenamefont {Gopakumar}\ and\ \citenamefont
  {Sinha}(2018)}]{Gopakumar:2018xqi}%
  \BibitemOpen
  \bibfield  {author} {\bibinfo {author} {\bibfnamefont {R.}~\bibnamefont
  {Gopakumar}}\ and\ \bibinfo {author} {\bibfnamefont {A.}~\bibnamefont
  {Sinha}},\ }\href {\doibase 10.1007/JHEP12(2018)040} {\bibfield  {journal}
  {\bibinfo  {journal} {JHEP}\ }\textbf {\bibinfo {volume} {12}},\ \bibinfo
  {pages} {040} (\bibinfo {year} {2018})},\ \Eprint
  {http://arxiv.org/abs/1809.10975} {arXiv:1809.10975 [hep-th]} \BibitemShut
  {NoStop}%
\bibitem [{\citenamefont {Polyakov}(1974)}]{Polyakov:1974gs}%
  \BibitemOpen
  \bibfield  {author} {\bibinfo {author} {\bibfnamefont {A.~M.}\ \bibnamefont
  {Polyakov}},\ }\href@noop {} {\bibfield  {journal} {\bibinfo  {journal} {Zh.
  Eksp. Teor. Fiz.}\ }\textbf {\bibinfo {volume} {66}},\ \bibinfo {pages} {23}
  (\bibinfo {year} {1974})}\BibitemShut {NoStop}%
\bibitem [{\citenamefont {Paulos}\ \emph {et~al.}(2017)\citenamefont {Paulos},
  \citenamefont {Penedones}, \citenamefont {Toledo}, \citenamefont {van Rees},\
  and\ \citenamefont {Vieira}}]{Paulos:2016fap}%
  \BibitemOpen
  \bibfield  {author} {\bibinfo {author} {\bibfnamefont {M.~F.}\ \bibnamefont
  {Paulos}}, \bibinfo {author} {\bibfnamefont {J.}~\bibnamefont {Penedones}},
  \bibinfo {author} {\bibfnamefont {J.}~\bibnamefont {Toledo}}, \bibinfo
  {author} {\bibfnamefont {B.~C.}\ \bibnamefont {van Rees}}, \ and\ \bibinfo
  {author} {\bibfnamefont {P.}~\bibnamefont {Vieira}},\ }\href {\doibase
  10.1007/JHEP11(2017)133} {\bibfield  {journal} {\bibinfo  {journal} {JHEP}\
  }\textbf {\bibinfo {volume} {11}},\ \bibinfo {pages} {133} (\bibinfo {year}
  {2017})},\ \Eprint {http://arxiv.org/abs/1607.06109} {arXiv:1607.06109
  [hep-th]} \BibitemShut {NoStop}%
\bibitem [{\citenamefont {van Rees}\ and\ \citenamefont
  {Zhao}(2023{\natexlab{a}})}]{vanRees:2022zmr}%
  \BibitemOpen
  \bibfield  {author} {\bibinfo {author} {\bibfnamefont {B.~C.}\ \bibnamefont
  {van Rees}}\ and\ \bibinfo {author} {\bibfnamefont {X.}~\bibnamefont
  {Zhao}},\ }\href {\doibase 10.1103/PhysRevLett.130.191601} {\bibfield
  {journal} {\bibinfo  {journal} {Phys. Rev. Lett.}\ }\textbf {\bibinfo
  {volume} {130}},\ \bibinfo {pages} {191601} (\bibinfo {year}
  {2023}{\natexlab{a}})},\ \Eprint {http://arxiv.org/abs/2210.15683}
  {arXiv:2210.15683 [hep-th]} \BibitemShut {NoStop}%
\bibitem [{\citenamefont {van Rees}\ and\ \citenamefont
  {Zhao}(2023{\natexlab{b}})}]{vanRees:2023fcf}%
  \BibitemOpen
  \bibfield  {author} {\bibinfo {author} {\bibfnamefont {B.~C.}\ \bibnamefont
  {van Rees}}\ and\ \bibinfo {author} {\bibfnamefont {X.}~\bibnamefont
  {Zhao}},\ }\href@noop {} {\  (\bibinfo {year} {2023}{\natexlab{b}})},\
  \Eprint {http://arxiv.org/abs/2312.02273} {arXiv:2312.02273 [hep-th]}
  \BibitemShut {NoStop}%
\bibitem [{\citenamefont {Hogervorst}\ and\ \citenamefont {van
  Rees}(2017)}]{Hogervorst:2017sfd}%
  \BibitemOpen
  \bibfield  {author} {\bibinfo {author} {\bibfnamefont {M.}~\bibnamefont
  {Hogervorst}}\ and\ \bibinfo {author} {\bibfnamefont {B.~C.}\ \bibnamefont
  {van Rees}},\ }\href {\doibase 10.1007/JHEP11(2017)193} {\bibfield  {journal}
  {\bibinfo  {journal} {JHEP}\ }\textbf {\bibinfo {volume} {11}},\ \bibinfo
  {pages} {193} (\bibinfo {year} {2017})},\ \Eprint
  {http://arxiv.org/abs/1702.08471} {arXiv:1702.08471 [hep-th]} \BibitemShut
  {NoStop}%
\bibitem [{\citenamefont {Carmi}\ \emph {et~al.}(2025)\citenamefont {Carmi},
  \citenamefont {Moreno},\ and\ \citenamefont {Sukholuski}}]{Carmi:2025mtx}%
  \BibitemOpen
  \bibfield  {author} {\bibinfo {author} {\bibfnamefont {D.}~\bibnamefont
  {Carmi}}, \bibinfo {author} {\bibfnamefont {J.}~\bibnamefont {Moreno}}, \
  and\ \bibinfo {author} {\bibfnamefont {S.}~\bibnamefont {Sukholuski}},\
  }\href {\doibase 10.1103/s5sx-9yjz} {\bibfield  {journal} {\bibinfo
  {journal} {Phys. Rev. D}\ }\textbf {\bibinfo {volume} {112}},\ \bibinfo
  {pages} {L101701} (\bibinfo {year} {2025})},\ \Eprint
  {http://arxiv.org/abs/2503.08774} {arXiv:2503.08774 [hep-th]} \BibitemShut
  {NoStop}%
\bibitem [{\citenamefont {Simmons-Duffin}\ \emph {et~al.}(2018)\citenamefont
  {Simmons-Duffin}, \citenamefont {Stanford},\ and\ \citenamefont
  {Witten}}]{Simmons-Duffin:2017nub}%
  \BibitemOpen
  \bibfield  {author} {\bibinfo {author} {\bibfnamefont {D.}~\bibnamefont
  {Simmons-Duffin}}, \bibinfo {author} {\bibfnamefont {D.}~\bibnamefont
  {Stanford}}, \ and\ \bibinfo {author} {\bibfnamefont {E.}~\bibnamefont
  {Witten}},\ }\href {\doibase 10.1007/JHEP07(2018)085} {\bibfield  {journal}
  {\bibinfo  {journal} {JHEP}\ }\textbf {\bibinfo {volume} {07}},\ \bibinfo
  {pages} {085} (\bibinfo {year} {2018})},\ \Eprint
  {http://arxiv.org/abs/1711.03816} {arXiv:1711.03816 [hep-th]} \BibitemShut
  {NoStop}%
\bibitem [{\citenamefont {Hogervorst}\ and\ \citenamefont
  {Rychkov}(2013)}]{Hogervorst:2013sma}%
  \BibitemOpen
  \bibfield  {author} {\bibinfo {author} {\bibfnamefont {M.}~\bibnamefont
  {Hogervorst}}\ and\ \bibinfo {author} {\bibfnamefont {S.}~\bibnamefont
  {Rychkov}},\ }\href {\doibase 10.1103/PhysRevD.87.106004} {\bibfield
  {journal} {\bibinfo  {journal} {Phys. Rev. D}\ }\textbf {\bibinfo {volume}
  {87}},\ \bibinfo {pages} {106004} (\bibinfo {year} {2013})},\ \Eprint
  {http://arxiv.org/abs/1303.1111} {arXiv:1303.1111 [hep-th]} \BibitemShut
  {NoStop}%
\bibitem [{\citenamefont {Komatsu}\ \emph {et~al.}(2020)\citenamefont
  {Komatsu}, \citenamefont {Paulos}, \citenamefont {Van~Rees},\ and\
  \citenamefont {Zhao}}]{Komatsu:2020sag}%
  \BibitemOpen
  \bibfield  {author} {\bibinfo {author} {\bibfnamefont {S.}~\bibnamefont
  {Komatsu}}, \bibinfo {author} {\bibfnamefont {M.~F.}\ \bibnamefont {Paulos}},
  \bibinfo {author} {\bibfnamefont {B.~C.}\ \bibnamefont {Van~Rees}}, \ and\
  \bibinfo {author} {\bibfnamefont {X.}~\bibnamefont {Zhao}},\ }\href {\doibase
  10.1007/JHEP11(2020)046} {\bibfield  {journal} {\bibinfo  {journal} {JHEP}\
  }\textbf {\bibinfo {volume} {11}},\ \bibinfo {pages} {046} (\bibinfo {year}
  {2020})},\ \Eprint {http://arxiv.org/abs/2007.13745} {arXiv:2007.13745
  [hep-th]} \BibitemShut {NoStop}%
\bibitem [{\citenamefont {Kravchuk}\ \emph {et~al.}(2020)\citenamefont
  {Kravchuk}, \citenamefont {Qiao},\ and\ \citenamefont
  {Rychkov}}]{Kravchuk:2020scc}%
  \BibitemOpen
  \bibfield  {author} {\bibinfo {author} {\bibfnamefont {P.}~\bibnamefont
  {Kravchuk}}, \bibinfo {author} {\bibfnamefont {J.}~\bibnamefont {Qiao}}, \
  and\ \bibinfo {author} {\bibfnamefont {S.}~\bibnamefont {Rychkov}},\ }\href
  {\doibase 10.1007/JHEP05(2020)137} {\bibfield  {journal} {\bibinfo  {journal}
  {JHEP}\ }\textbf {\bibinfo {volume} {05}},\ \bibinfo {pages} {137} (\bibinfo
  {year} {2020})},\ \Eprint {http://arxiv.org/abs/2001.08778} {arXiv:2001.08778
  [hep-th]} \BibitemShut {NoStop}%
\end{thebibliography}%

\end{document}